\documentclass[aps,pra,twocolumn,secnumarabic,superscriptaddress,longbibliography]{revtex4-2}
\usepackage{graphicx}
\usepackage{subfigure}
\usepackage{amsthm}
\usepackage{tensor}
\usepackage{color}
\usepackage[all]{xy}
\usepackage{tikz}
\usepackage{dsfont}
\usepackage[T1]{fontenc}
\usepackage{times,txfonts}
\usetikzlibrary{positioning}
\usepackage{braket}

\newtheorem{Thm}{Theorem}
\newtheorem{Lem}{Lemma}
\newtheorem{Prop}{Proposition}
\newtheorem{Cor}{Corollary}
\theoremstyle{definition}
\newtheorem{Def}{Definition}

\newcommand{\Tr}{\mathop{\mathrm{Tr}}\nolimits}

\begin{document}

\title{An operational genuine multipartite entanglement measure incorporating the Kempe invariant}

\author{Jeonghyeon Shin} 
\email{jh.shin234@gmail.com}
\affiliation{Center for Quantum Technology, Korea Institute of Science and Technology (KIST), Seoul 02792, Republic of Korea}
\affiliation{Department of Mathematics and Research Institute for Basic Sciences, Kyung Hee University, Seoul 02447, Republic of Korea}

\author{Minjin Choi}
\affiliation{Center for Quantum Information R\&D, Korea Institute of Science and Technology Information~(KISTI), Daejeon 34141, Republic of Korea}

\author{Jaehak Lee} 
\affiliation{Center for Quantum Technology, Korea Institute of Science and Technology (KIST), Seoul 02792, Republic of Korea}
\affiliation{Division of Quantum Information, KIST School, Korea University of Science and Technology (UST), Seoul 02792, Republic of Korea}

\author{Seung-Woo Lee}
\affiliation{Department of Physics, Pohang University of Science and Technology (POSTECH), Pohang 37673, Republic of Korea}

\author{Soojoon Lee}
\email{level@khu.ac.kr}
\affiliation{Department of Mathematics and Research Institute for Basic Sciences, Kyung Hee University, Seoul 02447, Republic of Korea}

\date{\today}

\begin{abstract}
We construct a measure of genuine multipartite entanglement (GME) for three-qubit states based on control power in quantum teleportation, which quantifies the fidelity gain provided by a third-party controller. For pure states, the measure is the geometric mean of the normalized fidelity gains for the three choices of controller. We extend it to mixed states by the convex roof and prove that it is a faithful GME monotone. We further show that, for genuinely entangled pure states with fixed pairwise concurrences and three-tangle, the measure increases strictly with the Kempe invariant, whereas measures determined solely by the pairwise concurrences and three-tangle remain constant. Under these constraints, a larger Kempe invariant signifies a lower unassisted teleportation fidelity for every pair and hence a larger gain from controller assistance, since each pair's optimal assisted fidelity remains fixed. Our construction thus connects the Kempe invariant, GME quantification, and a multipartite communication task through the benefit of controller assistance, giving the Kempe-dependent ordering a direct operational interpretation.
\end{abstract}
\maketitle

\section{Introduction}
\label{sec: Introduction}

Multipartite entanglement is a resource for quantum communication and quantum metrology \cite{Hillery1999,Epping2017,Toth2012}. A state exhibits genuine multipartite entanglement (GME) when its entanglement extends across all parties in the sense that the state cannot be expressed as a convex mixture of biseparable states \cite{Guhne2009}. Such states are used in quantum secret-sharing protocols \cite{Hillery1999} and can improve conference-key distribution rates in quantum networks \cite{Epping2017}. GME is also necessary to attain maximal sensitivity in certain metrological tasks \cite{Toth2012}. These applications motivate the quantification of GME for assessing and comparing multipartite resource states.

Three fundamental axioms have been proposed as essential requirements for a consistent quantification of GME \cite{Li2022,Ma2025}. A GME measure must be nonincreasing on average under local operations and classical communication (LOCC) \cite{Vidal2000,Plenio2007}, vanish on biseparable states, and take a positive value on every genuinely multipartite entangled state \cite{Ma2011}. For pure states, this distinction can be made through entanglement across bipartitions. Several GME measures use this structure. The GME concurrence takes the minimum concurrence over these cuts \cite{Ma2011}, the geometric mean of bipartite concurrences combines their contributions \cite{Li2022}, and the generalized geometric measure uses the largest Schmidt coefficient over all cuts \cite{Sen2010}. A different approach defines GME measures directly from performance in a multipartite task, as in measures based on the fidelity achievable in controlled teleportation \cite{Choi2023}.

A complementary approach to characterizing three-qubit entanglement is based on local-unitary (LU) invariants. Three qubit pure states have five continuous degrees of freedom up to local unitaries \cite{Sudbery2001}, which can be parametrized generically by the three pairwise concurrences \cite{Wootters1998}, the three-tangle \cite{Coffman2000}, and the Kempe invariant \cite{Kempe1999}. The pairwise concurrences and three-tangle determine the single-qubit spectra through the Coffman--Kundu--Wootters relations and hence fix the bipartition-based measures discussed above \cite{Coffman2000}. They also determine the optimal assisted teleportation fidelities and the associated teleportation-capability measures \cite{Lee2005,Choi2023}. These measures therefore remain unchanged when the Kempe invariant varies at fixed concurrence and tangle data. Beyond these measures, the role of the Kempe invariant in entanglement monotones and its interpretation in terms of observables have been studied \cite{Szalay2026,Williamson2011}. In particular, the existence of an entanglement monotone depending on this invariant follows from the fact that the complete set of pure-state monotones distinguishes LU orbits \cite{Gingrich2002}. An entanglement monotone that depends on the Kempe invariant was previously proposed, but it fails the faithfulness requirement for GME measures \cite{Oreshkov2006,Oreshkov2007}. It is also known that the Kempe invariant is proportional to the average Hilbert--Schmidt inner product between a local observable and its image after a transformation around a closed path through the three qubits \cite{Williamson2011}. This raises the question of whether a faithful GME measure can link the ordering of states by the Kempe invariant to a performance gain in a multipartite communication task.

In this work, we construct a GME measure for three-qubit states from control power, the fidelity gain supplied by the controller in controlled teleportation \cite{Li2014,Jeong2016}. Each control power compares the optimized assisted fidelity with the unassisted standard-teleportation fidelity for the corresponding pair. For pure state, we combine the three normalized control powers through their geometric mean to account for the contribution of every party as controller and extend the measure to mixed state via convex roof. We prove that the resulting quantity vanishes exactly on biseparable states and is nonincreasing on average under LOCC. To prove this monotonicity, we show that for every two-qubit state, any decrease in the average fully entangled fraction under a local measurement on either qubit is bounded by half the corresponding decrease in average concurrence of assistance. We also determine the pure-state maxima within the GHZ and W classes, attained uniquely up to local unitaries by their standard representatives.

We further prove that, among genuinely entangled three-qubit pure states with fixed pairwise concurrences and three-tangle, our measure increases strictly with the Kempe invariant. Under these constraints, a larger Kempe invariant corresponds to a strictly lower unassisted standard-teleportation fidelity for each pair. Since the optimal assisted fidelity of each pair is fixed, the gain is strictly larger for every choice of controller. We illustrate this ordering with a continuous family of states for which the proposed measure varies although the GME concurrence and the existing teleportation-capability measures remain constant. Theorefore, our measure establishes an operational connection between GME and the Kempe invariant through the fidelity gain in a concrete multipartite communication task.

\section{Controlled teleportation and control power}
\label{sec:controlled-teleportation}

Quantum teleportation enables the transfer of an unknown qubit state between two parties using local operations and classical communication with a shared two-qubit state $\rho_{AB}$ \cite{Bennett1993}. The performance of a teleportation scheme with channel $\Lambda_{\rho_{AB}}$ is characterized by its average fidelity \cite{Horodecki1999},
\begin{equation}
F(\Lambda_{\rho_{AB}})=\int d\xi\,\bra{\xi}\Lambda_{\rho_{AB}}(\ket{\xi}\bra{\xi})\ket{\xi},
\label{eq:average-teleportation-fidelity}
\end{equation}
where the integral averages over all one-qubit pure states according to the uniform distribution $d\xi$. To characterize the shared state $\rho_{AB}$, we introduce the fully entangled fraction (FEF), defined by
\begin{equation}
f(\rho_{AB})=\max_{\ket{e}_{AB}}\bra{e}\rho_{AB}\ket{e}_{AB},
\label{eq:fully-entangled-fraction}
\end{equation}
where the maximum is taken over maximally entangled two-qubit states $\ket{e}_{AB}$. It has been shown that the maximal average fidelity of standard teleportation, optimized over local unitaries, is linearly related to the FEF \cite{Horodecki1999,Badziag2000},
\begin{equation}
F_{\mathrm{std}}(\rho_{AB})=\frac{2f(\rho_{AB})+1}{3}.
\label{eq:standard-teleportation}
\end{equation}
The given state $\rho_{AB}$ is said to be useful for standard teleportation if and only if $F_{\mathrm{std}}(\rho_{AB})>2/3$, so that its fidelity exceeds the optimal classical value \cite{Massar1995,Horodecki1999}.

Controlled teleportation extends this protocol to three parties sharing a three-qubit pure state $\ket{\psi}_{ABC}$, with $X$ acting as the controller and $Y$ and $Z$ as the sender and receiver, respectively, where $\{X,Y,Z\}=\{A,B,C\}$. Let $\rho_X=\Tr_{YZ}\ket{\psi}\bra{\psi}$ and $\rho_{YZ}=\Tr_X\ket{\psi}\bra{\psi}$ be the reduced states. The controller applies a unitary $U_X$ to its qubit, measures in the computational basis, and communicates the outcome $t$ to $Y$ and $Z$, who use the conditional state $\rho_{YZ}^{(t)}$ for standard teleportation. As in the two-qubit case, the maximal average fidelity with controller assistance is linearly related to the maximal average FEF of the conditional states on $YZ$ \cite{Lee2005,Jeong2016},
\begin{equation}
F_{\mathrm{CT}}^X(\ket{\psi})
=\frac{2f_{\mathrm{CT}}^X(\ket{\psi})+1}{3},
\label{eq:controlled-fidelity-fef}
\end{equation}
where
\[
f_{\mathrm{CT}}^X(\ket{\psi})
=\max_{U_X}\sum_{t=0}^{1}
\bra{t}U_X\rho_XU_X^\dagger\ket{t}_X
f(\rho_{YZ}^{(t)}),
\]
and the maximum is taken over $2\times2$ unitary matrices $U_X$. 

For a three-qubit pure state $\ket{\psi}$, the partial tangle of the sender--receiver pair determines the maximal controlled fidelity \cite{Lee2005},
\begin{equation}
F_{\mathrm{CT}}^X(\ket{\psi})=\frac{2+\tau_{YZ}}{3},
\label{eq:controlled-fidelity-partial-tangle}
\end{equation}
where
\begin{equation}
\tau_{YZ}=\sqrt{C_{YZ}^2+\tau}
\label{eq:partial-tangle}
\end{equation}
is the partial tangle and $C_{YZ}=C(\rho_{YZ})$ denotes the two-qubit concurrence \cite{Wootters1998}. The three-tangle $\tau$ satisfies the Coffman--Kundu--Wootters (CKW) relation \cite{Coffman2000}
\begin{equation}
C_{X(YZ)}^2=C_{XY}^2+C_{XZ}^2+\tau,
\label{eq:ckw}
\end{equation}
with $C_{X(YZ)}=2\sqrt{\det\rho_X}$ and $\tau$ independent of the choice of $X$. The pairwise concurrences quantify two-qubit entanglement, whereas the three-tangle quantifies the residual three-party entanglement in the CKW relation.

The partial tangle $\tau_{YZ}$ also equals the concurrence of assistance (CoA) of the reduced state $\rho_{YZ}$ for three qubit pure state \cite{Chi2010}. The CoA of a two-qubit state $\rho_{AB}$ is defined by \cite{Laustsen2003}
\[
C_a(\rho_{AB})=\max_{\{q_i,\ket{\eta_i}_{AB}\}}
\sum_i q_i C(\ket{\eta_i}_{AB}),
\]
where the maximization is over ensembles satisfying $\rho_{AB}=\sum_i q_i\ket{\eta_i}_{AB}\bra{\eta_i}$. By definition, the CoA is a concave function of $\rho_{AB}$. For a reduced state $\rho_{YZ}$ of the three-qubit state $\ket{\psi}$, it satisfies 
\begin{equation}
C_a(\rho_{YZ})^2=C_{YZ}^2+\tau.
\label{eq:coa-three-tangle}
\end{equation}
The controlled fidelity in Eq.~(\ref{eq:controlled-fidelity-partial-tangle}) can therefore be written as
\begin{equation}
F_{\mathrm{CT}}^X(\ket{\psi})=\frac{2+C_a(\rho_{YZ})}{3}.
\label{eq:optimized-controlled-fidelity}
\end{equation}

Since $F_{\mathrm{CT}}^X(\ket{\psi})\ge F_{\mathrm{std}}(\rho_{YZ})$, the control power of party $X$ is defined as the nonnegative difference between the teleportation fidelities with and without its assistance \cite{Jeong2016},
\begin{equation}
P_X(\ket{\psi})=F_{\mathrm{CT}}^X(\ket{\psi})-F_{\mathrm{std}}(\rho_{YZ}).
\label{eq:control-power-definition}
\end{equation}
For pure three-qubit states, the control power satisfies $P_X\le1/3$ \cite{Li2014}.
We use the normalized control power $\mathcal P_X(\ket{\psi}):=3P_X(\ket{\psi})$. Substituting Eqs.~(\ref{eq:standard-teleportation}) and (\ref{eq:optimized-controlled-fidelity}) gives
\begin{equation}
\begin{array}{r@{}l}
\mathcal P_X(\ket{\psi})
&\displaystyle{}=1+C_a(\rho_{YZ})-2f(\rho_{YZ})\\[3pt]
&\displaystyle{}=1+\sqrt{C_{YZ}^2+\tau}-2f(\rho_{YZ}).
\end{array}
\label{eq:directional-control-power}
\end{equation}
The quantities $\mathcal P_A$, $\mathcal P_B$, and $\mathcal P_C$ correspond to the pairs $BC$, $AC$, and $AB$, respectively, with each controller's measurement optimized separately.

\section{A control-power-based GME measure}
\label{sec:gcp-measure}

The normalized control power introduced in the previous section quantifies the improvement in teleportation fidelity provided by a controller. For a pure three-qubit state that is product across the bipartition $X|YZ$, a measurement by $X$ cannot change the state shared by $Y$ and $Z$, and hence $\mathcal P_X=0$. Thus, a measure of genuine tripartite entanglement constructed from control power should vanish whenever any one of the three parties is uncorrelated with the others. We combine the three control powers through their geometric mean, which satisfies this requirement and is symmetric under permutations of the parties.

\begin{Def}
\label{def:gcp-measure}
For a three-qubit pure state $\ket{\psi}$, define
\begin{equation}
e_{\mathrm{GCP}}(\ket{\psi})
:=\left[\prod_{X=A,B,C}\mathcal P_X(\ket{\psi})\right]^{1/3}.
\label{eq:gcp-pure-definition}
\end{equation}
For mixed three-qubit states, the measure is defined by the convex-roof extension \cite{Vidal2000,Uhlmann2010},
\begin{equation}
e_{\mathrm{GCP}}(\rho)
=\min_{{q_i,|\psi_i\rangle}}
\sum_i q_i e_{\mathrm{GCP}}(|\psi_i\rangle),
\end{equation}
where the minimum is over all pure-state decompositions of $\rho$.
\end{Def}

By construction, the proposed measure is nonnegative and bounded by unity. It is also invariant under local unitary operations and permutations of the parties, since the concurrence of assistance and the fully entangled fraction are locally unitary invariant. To establish that it is a GME measure, it remains to prove faithfulness and average monotonicity under LOCC. The convex-roof extension is convex by definition and preserves these properties once they are established for pure states \cite{Vidal2000}. It is therefore sufficient to consider pure states.

To prove monotonicity under local operations by the sender or receiver, we use the following bound.

\begin{Lem}
\label{lem:gcp-filtering}
Let $\rho_{YZ}$ be a two-qubit state and let $\{K_t\}$ be a finite family of $2\times2$ Kraus operators on $Y$ satisfying $\sum_tK_t^\dagger K_t=I$. Set $\sigma_t=(K_t\otimes I)\rho_{YZ}(K_t^\dagger\otimes I)$, $p_t=\Tr\sigma_t$, $\rho_{YZ}^{(t)}=\sigma_t/p_t$ for $p_t>0$, and $s=\sum_t|\det K_t|$. Then $s\le1$ and
\begin{equation}
\begin{array}{r@{}l}
\displaystyle\sum_t p_t C_a(\rho_{YZ}^{(t)})
&{}=sC_a(\rho_{YZ}),\\[3pt]
\displaystyle\sum_t p_t f(\rho_{YZ}^{(t)})
&\displaystyle{}\ge f(\rho_{YZ})-\frac{1-s}{2}C_a(\rho_{YZ}).
\end{array}
\label{eq:gcp-filtering}
\end{equation}
The same relations hold for an instrument on $Z$.
\end{Lem}

The proof is given in Appendix~\ref{app:fef-filtering}. We use these relations to establish strong monotonicity of $e_{\mathrm{GCP}}$ on pure states.

\begin{Lem}
\label{lem:gcp-monotonicity}
For three-qubit pure states, $e_{\mathrm{GCP}}$ does not increase on average under finite-round LOCC.
\end{Lem}

\begin{proof}
Each local step of an LOCC protocol can be represented by a positive-operator-valued measure (POVM) with elements $\{K_t^\dagger K_t\}$, where the Kraus operators $\{K_t\}$ satisfy $\sum_tK_t^\dagger K_t=I$. Let $p_t$ denote the outcome probabilities and $\ket{\psi_t}$ the outputs for outcome $t$. If the operation acts on the controller $X$, the average pair state satisfies $\rho_{YZ}=\sum_t p_t\rho_{YZ}^{(t)}$, where $\rho_{YZ}^{(t)}=\Tr_X\ket{\psi_t}\bra{\psi_t}$ is the reduced state of $YZ$ conditioned on outcome $t$. Concavity of $C_a$ and convexity of $f$ therefore give
\[
\sum_t p_t[1+C_a(\rho_{YZ}^{(t)})-2f(\rho_{YZ}^{(t)})]
\le1+C_a(\rho_{YZ})-2f(\rho_{YZ}).
\]

For an operation on $Y$ or $Z$, substituting Eq.~(\ref{eq:gcp-filtering}) into $\mathcal P_X=1+C_a(\rho_{YZ})-2f(\rho_{YZ})$ gives the same average bound, since $sC_a+(1-s)C_a=C_a$. Thus, an instrument on any party satisfies
\begin{equation}
\sum_t p_t\mathcal P_X(\ket{\psi_t})
\le\mathcal P_X(\ket{\psi}),\qquad X=A,B,C.
\label{eq:gcp-directional-monotonicity}
\end{equation}

Iterating Eq.~(\ref{eq:gcp-directional-monotonicity}) over a finite-round LOCC protocol gives $\sum_m p_m\mathcal P_X(\ket{\psi_m})\le\mathcal P_X(\ket\psi)$ for each controller $X$, where $\{p_m,\ket{\psi_m}\}$ is the output ensemble. Applying the generalized H\"older inequality to the geometric mean gives
\begin{equation}
\begin{array}{r@{}l}
\displaystyle\sum_m p_m e_{\mathrm{GCP}}(\ket{\psi_m})
&\displaystyle{}\le\prod_X\left[\sum_m p_m\mathcal P_X(\ket{\psi_m})\right]^{1/3}\\[5pt]
&{}\le e_{\mathrm{GCP}}(\ket{\psi}).
\end{array}
\label{eq:gcp-pure-monotonicity}
\end{equation}
\end{proof}

To establish faithfulness, it remains to identify the zero set of $e_{\mathrm{GCP}}$.

\begin{Lem}
\label{lem:gcp-faithfulness}
For every three-qubit pure state $\ket{\psi}$, $e_{\mathrm{GCP}}(\ket{\psi})=0$ if and only if $\ket{\psi}$ is biseparable.
\end{Lem}

\begin{proof}
If a pure state is biseparable, it is product across at least one bipartition. Up to relabeling the parties, consider the partition $X|YZ$, for which $\ket{\psi}=\ket{\alpha}_X\otimes\ket{\phi}_{YZ}$. A measurement by the controller $X$ then leaves $\ket{\phi}_{YZ}$ unchanged, so $\mathcal P_X=0$ and $e_{\mathrm{GCP}}=0$.

Conversely, every genuinely multipartite entangled three-qubit pure state belongs to one of two classes under stochastic LOCC (SLOCC) \cite{Dur2000}. A state in either class can be converted with nonzero probability to its standard representative, defined by
\[
\ket{\mathrm{GHZ}}=\frac{\ket{000}+\ket{111}}{\sqrt{2}},
\qquad
\ket{W}=\frac{\ket{100}+\ket{010}+\ket{001}}{\sqrt{3}}.
\]
For these representatives, the reduced two-qubit states have $(C_a,f)=(1,1/2)$ and $(2/3,2/3)$, respectively, giving $e_{\mathrm{GCP}}(\ket{\mathrm{GHZ}})=1$ and $e_{\mathrm{GCP}}(\ket{W})=1/3$~\footnote{For the standard W state, $f(\rho_{YZ})=2/3$, giving $F_{\mathrm{std}}(\rho_{YZ})=7/9$ and $F_{\mathrm{CT}}^X=8/9$ for every controller $X$. Hence the unnormalized control power is $P_X=1/9$, rather than $2/9$ as reported in Ref.~\cite{Jeong2016}. With our normalization, this gives $\mathcal P_X=1/3$.}.

If the successful branch has probability $p>0$ and output $\ket{\chi}$, Eq.~(\ref{eq:gcp-pure-monotonicity}) and nonnegativity give
\[
e_{\mathrm{GCP}}(\ket{\psi})
\ge\sum_m p_m e_{\mathrm{GCP}}(\ket{\psi_m})
\ge p\,e_{\mathrm{GCP}}(\ket{\chi})>0.
\]

\end{proof}

The convex roof preserves faithfulness and strong monotonicity and is convex by construction. Together with local unitary invariance, these properties give the following result.

\begin{Thm}[Control-power-based GME measure]
\label{thm:gcp-measure}
The quantity $e_{\mathrm{GCP}}$ is a GME measure on three-qubit states.
\end{Thm}

\begin{proof}
For a mixed state $\rho$, choose a pure-state decomposition $\rho=\sum_i q_i\ket{\psi_i}\bra{\psi_i}$ that attains the minimum defining $e_{\mathrm{GCP}}(\rho)$. Such a decomposition exists by continuity and compactness \cite{Uhlmann2010}. If $e_{\mathrm{GCP}}(\rho)=0$, each term with $q_i>0$ has $e_{\mathrm{GCP}}(\ket{\psi_i})=0$ by nonnegativity. Lemma~\ref{lem:gcp-faithfulness} then implies that each such component is biseparable, so $\rho$ is biseparable. Conversely, any biseparable state admits a decomposition into biseparable pure states with zero average, so its convex-roof minimum is zero.

Suppose one party performs a local measurement with Kraus operators $\{K_t\}$ satisfying $\sum_t K_t^\dagger K_t=I$, with identities on the other parties implicit. Set $r_{it}=\|K_t\ket{\psi_i}\|^2$ and $\ket{\psi_{it}}=K_t\ket{\psi_i}/\sqrt{r_{it}}$ for $r_{it}>0$. Outcome $t$ has probability $p_t=\sum_i q_i r_{it}$ and conditional state
\[
\rho_t=\sum_i\frac{q_i r_{it}}{p_t}\ket{\psi_{it}}\bra{\psi_{it}},\qquad p_t>0.
\]

The convex-roof definition and Eq.~(\ref{eq:gcp-pure-monotonicity}) give
\[
\begin{array}{r@{}l}
\displaystyle\sum_t p_t e_{\mathrm{GCP}}(\rho_t)
&\displaystyle{}\le\sum_{i,t}q_i r_{it}e_{\mathrm{GCP}}(\ket{\psi_{it}})\\[4pt]
&\displaystyle{}\le\sum_i q_i e_{\mathrm{GCP}}(\ket{\psi_i})
=e_{\mathrm{GCP}}(\rho),
\end{array}
\]
which completes the proof.
\end{proof}

The geometric mean combines the control powers of all three parties. Since each control power is nonincreasing on average under LOCC, we can also use the smallest of the three, which quantifies the fidelity gain available from the least effective controller.

\begin{Cor}[Minimal control power]
\label{cor:minimal-control-power}
For pure three-qubit states, let $e_{\min}(\ket\psi)=\min_{X=A,B,C}\mathcal P_X(\ket\psi)$. This quantity and its convex-roof extension to mixed states define a faithful GME measure.
\end{Cor}
\begin{proof}
The bound for each controller in the proof of Lemma~\ref{lem:gcp-monotonicity} gives
\[
\begin{array}{r@{}l}
\displaystyle\sum_m p_m e_{\min}(\ket{\psi_m})
&\displaystyle{}\le\min_X\sum_m p_m\mathcal P_X(\ket{\psi_m})\\[3pt]
&{}\le\min_X\mathcal P_X(\ket\psi)=e_{\min}(\ket\psi).
\end{array}
\]
The minimum and geometric mean of three nonnegative factors have the same zero set, so Lemma~\ref{lem:gcp-faithfulness} proves pure-state faithfulness. The convex-roof argument of Theorem~\ref{thm:gcp-measure} then applies unchanged.
\end{proof}
For pure states, $e_{\min}$ is three times the minimal control power introduced in Ref.~\cite{Jeong2016}. Thus directional monotonicity establishes a GME measure for both choices of aggregation.

\section{Properties of our GME measure}
\label{sec:gcp-properties}

In this section, we examine the properties of $e_{\mathrm{GCP}}$ for three-qubit pure states. We first determine its ranges and maximizers in the GHZ and W classes and illustrate its behavior with representative state families. We then establish a strict ordering with respect to the Kempe invariant for GME states at fixed pairwise concurrences and three-tangle, and compare it with existing GME measures through explicit examples.

\subsection{Ranges and maximizers}
\label{sec:gcp-ranges}

Genuinely multipartite entangled three-qubit pure states fall into two inequivalent SLOCC classes, represented by the GHZ and W classes \cite{Dur2000}. The maximal values of $e_{\mathrm{GCP}}$ in these classes and their equality conditions are established below.

\begin{Prop}[Pure-state maxima]
\label{prop:gcp-maximizers}
For three-qubit pure states, $e_{\mathrm{GCP}}\le1$, with equality if and only if the state is LU equivalent to $\ket{\mathrm{GHZ}}$. Within the W SLOCC class, $e_{\mathrm{GCP}}\le1/3$, with equality if and only if the state is LU equivalent to $\ket{W}$.
\end{Prop}

\begin{proof}
The global bound follows from $0\le\mathcal P_X\le1$. If $e_{\mathrm{GCP}}=1$, all three factors must equal one. Choose a controller measurement attaining the maximal average FEF, with outcome probabilities $p_t$ and conditional states $\rho_{YZ}^{(t)}$. For each of its two outcomes, $\rho_{YZ}\ge p_t\rho_{YZ}^{(t)}$ implies $f(\rho_{YZ})\ge p_t f(\rho_{YZ}^{(t)})$. Summing these inequalities and using Eqs.~(\ref{eq:controlled-fidelity-fef}) and (\ref{eq:optimized-controlled-fidelity}) gives
\[
2f(\rho_{YZ})\ge\sum_{t=0}^{1}p_t f(\rho_{YZ}^{(t)})
=\frac{1+C_a(\rho_{YZ})}{2}.
\]
Thus $f(\rho_{YZ})\ge[1+C_a(\rho_{YZ})]/4$. Substitution into Eq.~(\ref{eq:directional-control-power}) gives $\mathcal P_X\le[1+C_a(\rho_{YZ})]/2\le1$. Equality therefore requires $C_a=1$ and $f=1/2$ for every pair. Equation~(\ref{eq:coa-three-tangle}) then gives $C_{AB}^2=C_{AC}^2=C_{BC}^2=1-\tau$. By the CKW relation, $C_{A(BC)}^2=2-\tau\le1$. Since $\tau\le1$, we obtain $\tau=1$, which characterizes the LU orbit of the standard GHZ state \cite{Acin2000}. Conversely, $e_{\mathrm{GCP}}(\ket{\mathrm{GHZ}})=1$, so LU invariance gives equality throughout this orbit.

Up to local unitaries, a three-qubit W-class pure
state can be written as \cite{Dur2000}
\[
\ket{\psi_W}_{ABC}=b\ket{000}+a\ket{100}+d\ket{010}+c\ket{001},
\]
where $a,c,d>0$, $b\ge0$, and $a^2+b^2+c^2+d^2=1$. Direct evaluation of the control power in Eq.~(\ref{eq:directional-control-power}) gives
\[
\begin{array}{l}
\displaystyle\mathcal P_A=\frac{1+2cd-\sqrt{\Delta_A}}2,\\[5pt]
\Delta_A=(1+2cd-2a^2)^2+4a^2b^2.
\end{array}
\]
Since $\sqrt{\Delta_A}\ge|1+2cd-2a^2|\ge1+2cd-2a^2$, we obtain $\mathcal P_A\le a^2$. Relabeling gives $\mathcal P_B\le d^2$ and $\mathcal P_C\le c^2$. The arithmetic--geometric mean (AM--GM) inequality then gives
\[
e_{\mathrm{GCP}}(\ket{\psi_W}_{ABC})
\le(a^2c^2d^2)^{1/3}
\le\frac{1-b^2}{3}\le\frac13.
\]
Equality at $1/3$ requires $b=0$ and $a=c=d=1/\sqrt3$. These coefficients give the standard W state, whose three control powers equal $1/3$. The conditions are therefore also sufficient.
\end{proof}

For the generalized GHZ states $\ket{G_\theta}=\cos\theta\ket{000}+\sin\theta\ket{111}$, with $0<\theta\le\pi/4$, each normalized control power equals $\sin2\theta$. Thus $e_{\mathrm{GCP}}=\sin2\theta$ spans the full GHZ-class range $(0,1]$.

\begin{figure}[!htbp]
\centering
\includegraphics[width=\columnwidth]{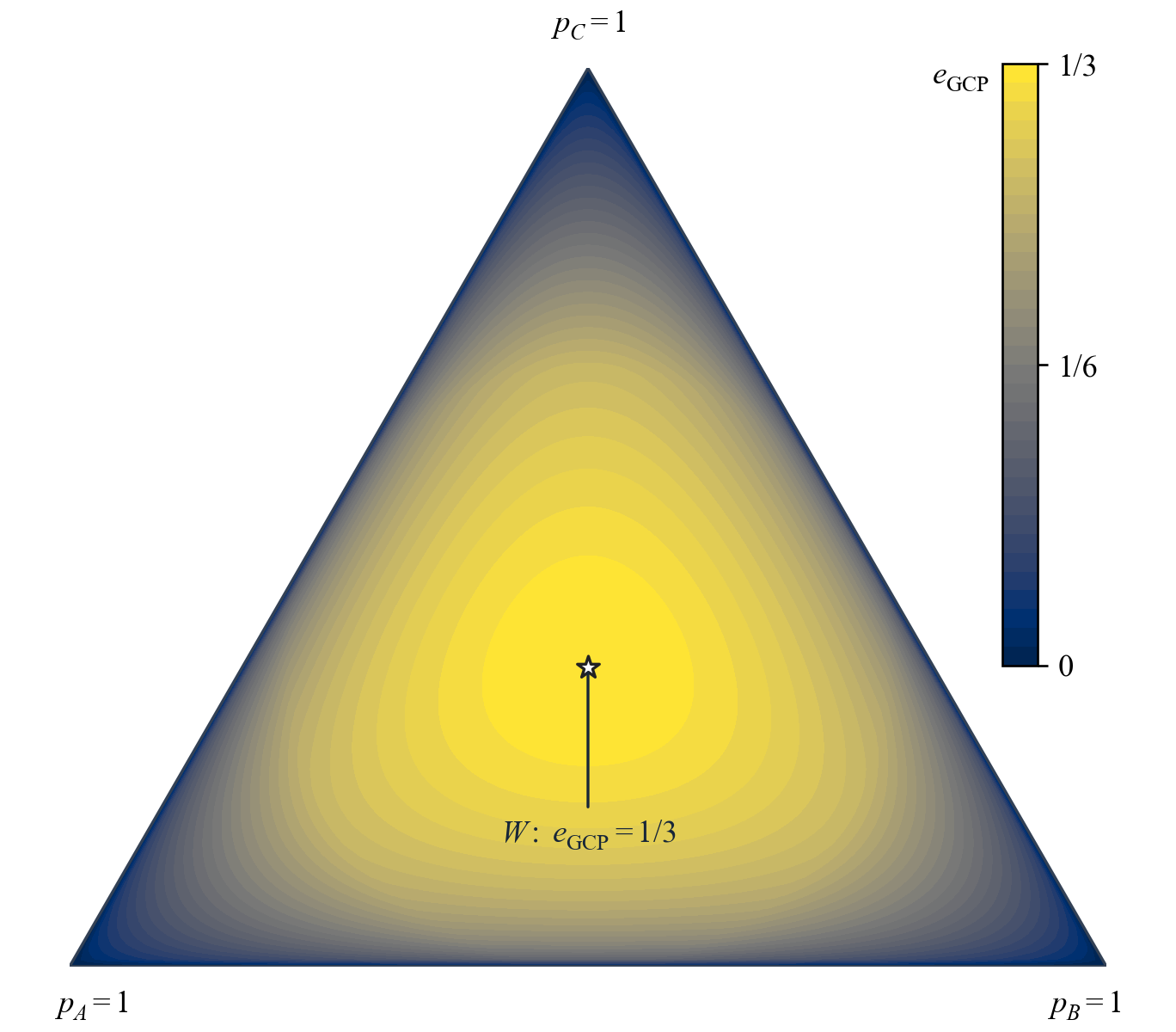}
\caption{The measure $e_{\mathrm{GCP}}$ for the single-excitation family in Eq.~(\ref{eq:w-simplex-family}), evaluated using Eq.~(\ref{eq:w-simplex-powers}). The center represents the standard W state and attains $1/3$. The edges are biseparable and have zero measure and the vertices are fully separable.}
\label{fig:w-simplex}
\end{figure}

The W-class range and its biseparable boundary are illustrated by the single-excitation family
\begin{equation}
\ket{W(\mathbf p)}
=\sqrt{p_A}\ket{100}+\sqrt{p_B}\ket{010}+\sqrt{p_C}\ket{001},
\label{eq:w-simplex-family}
\end{equation}
where $p_X\ge0$ and $p_A+p_B+p_C=1$. Setting $b=0$ in the expression for $\mathcal P_A$ in the proof of Proposition~\ref{prop:gcp-maximizers} and relabeling the qubits gives
\begin{equation}
\mathcal P_X=\min\{p_X,(\sqrt{p_Y}+\sqrt{p_Z})^2\}.
\label{eq:w-simplex-powers}
\end{equation}
Figure~\ref{fig:w-simplex} shows $e_{\mathrm{GCP}}$ as a function of the excitation distribution. The symmetric distribution attains the W-class maximum. On every edge, at least one directional factor vanishes and the states are biseparable. Choosing $p_A=q$ and $p_B=p_C=(1-q)/2$, with $0<q\le1/3$, gives $e_{\mathrm{GCP}}=[q(1-q)^2/4]^{1/3}$. This expression increases continuously from its limiting value zero to $1/3$.

\subsection{Dependence on the Kempe invariant}
\label{sec:kempe-dependence}

For generic pure three-qubit states, a convenient choice of five independent continuous LU invariants is $\{C_{AB},C_{AC},C_{BC},\tau,\kappa\}$ \cite{Sudbery2001}, where $C_{XY}$ is the concurrence of the reduced state of qubits $X$ and $Y$ \cite{Wootters1998}, $\tau$ is the three-tangle \cite{Coffman2000}, and $\kappa$ is the Kempe invariant \cite{Kempe1999}, which we write in the permutation-symmetric form \cite{Kempe1999,Sudbery2001}
\begin{equation}
\kappa=3\Tr[(\rho_B\otimes\rho_C)\rho_{BC}]
-\Tr\rho_B^3-\Tr\rho_C^3.
\label{eq:kempe-invariant}
\end{equation}
Complete LU classification also requires discrete information \cite{Acin2001}. Even when the three pairwise concurrences and the three-tangle are fixed, the Kempe invariant can distinguish states, capturing information not determined by these four quantities.

In the following theorem, we show $e_{\mathrm{GCP}}$ strictly orders pure GME states by $\kappa$ at fixed pairwise concurrences and three-tangle.

\begin{Thm}[Kempe-invariant ordering]
\label{thm:kempe-ordering}
Let $\ket{\psi_1}$ and $\ket{\psi_2}$ be genuinely multipartite entangled three-qubit pure states with identical $(C_{AB},C_{AC},C_{BC},\tau)$, and let $\kappa_i$ denote the Kempe invariant of $\ket{\psi_i}$ for $i=1,2$. Then $\kappa_2>\kappa_1$ if and only if
\begin{equation}
e_{\mathrm{GCP}}(\ket{\psi_2})>e_{\mathrm{GCP}}(\ket{\psi_1}).
\label{eq:kempe-gcp-ordering}
\end{equation}

\end{Thm}

\begin{proof}
We first prove that $\kappa_2>\kappa_1$ implies $e_{\mathrm{GCP}}(\ket{\psi_2})>e_{\mathrm{GCP}}(\ket{\psi_1})$ by comparing the three control powers entering their geometric means. The fixed pairwise concurrences and three-tangle also fix $C_a(\rho_{YZ}^{(i)})=\sqrt{C_{YZ}^2+\tau}$. Eq.~(\ref{eq:directional-control-power}) therefore gives
\[
\mathcal P_X(\ket{\psi_2})-\mathcal P_X(\ket{\psi_1})
=2\bigl[f(\rho_{YZ}^{(1)})-f(\rho_{YZ}^{(2)})\bigr].
\]
It thus suffices to show that $f(\rho_{YZ}^{(2)})<f(\rho_{YZ}^{(1)})$ for every pair.

For any two-qubit state $\rho$, let $\widetilde\rho=(\sigma_y\otimes\sigma_y)\rho^*(\sigma_y\otimes\sigma_y)$ denote its spin flip, where $\sigma_y$ is the Pauli $Y$ matrix and $*$ denotes complex conjugation in the computational basis. Set $R=(\rho+\widetilde\rho)/2$. Then the FEF equals the largest eigenvalue of $R$ \cite{Grondalski2002, Bennett1996},
\begin{equation}
f(\rho)=\lambda_{\max}(R).
\label{eq:fef-spinflip-spectrum}
\end{equation}
In the magic basis, a Bell basis with suitable phase choices, maximally entangled pure states are precisely those with real coefficients up to a global phase \cite{Bennett1996}, and spin flip acts as complex conjugation on density matrices \cite{Wootters1998}. Hence $R$ is represented by the real part of $\rho$. The imaginary part is antisymmetric and has zero expectation in any real vector. The FEF optimization therefore reduces to maximizing the expectation value of the real symmetric matrix representing $R$ over real unit vectors. Its largest eigenvalue has a real eigenvector, which attains this maximum.

Consider the pair $BC$ and write $\rho_i=\rho_{BC}^{(i)}$ and $R_i=(\rho_i+\widetilde\rho_i)/2$. Define the local purities $\alpha=\Tr\rho_A^2$, $\beta=\Tr\rho_B^2$, and $\gamma=\Tr\rho_C^2$. The CKW relation gives $\alpha=1-(C_{AB}^2+C_{AC}^2+\tau)/2$, and similarly for $\beta$ and $\gamma$, so these purities are shared by the two states. The moments and determinant derived in Appendix~\ref{app:kempe-polynomial} give the characteristic polynomial
\begin{equation}
\begin{array}{r@{}l}
\chi_i(u)&\displaystyle{}=\det(uI_4-R_i)\\
&\displaystyle{}=u^4-u^3+\frac{1-2\alpha+\beta+\gamma}{4}u^2\\
&\displaystyle{}\quad+\left(\frac{\kappa_i}{6}+\frac1{12}-\frac{\beta+\gamma}{8}\right)u
+\frac{(\beta-\gamma)^2}{64}.
\end{array}
\label{eq:kempe-characteristic-polynomial}
\end{equation}
Only the linear coefficient depends on $\kappa_i$. Hence
\begin{equation}
\chi_2(u)-\chi_1(u)=\frac{\kappa_2-\kappa_1}{6}\,u.
\label{eq:kempe-polynomial-difference}
\end{equation}

Let $f_i=\lambda_{\max}(R_i)=f(\rho_i)$. For $\kappa_2>\kappa_1$, evaluating Eq.~(\ref{eq:kempe-polynomial-difference}) at $f_2$ yields
\[
\chi_1(f_2)=-\frac{\kappa_2-\kappa_1}{6}f_2<0.
\]
If $f_2\ge f_1$, every factor in $\chi_1(f_2)=\prod_{j=1}^4[f_2-\lambda_j(R_1)]$ would be nonnegative, a contradiction. Thus $f_2<f_1$, including repeated or zero eigenvalues. Relabeling the parties gives the same ordering for every pair.

Each control power therefore increases strictly. Their positivity for GME inputs gives strict increase of the geometric mean, proving the forward implication. For $\kappa_2=\kappa_1$, the characteristic polynomials coincide, so the FEFs, control powers, and $e_{\mathrm{GCP}}$ are equal. Interchanging the states gives the opposite ordering when $\kappa_2<\kappa_1$, completing the converse.
\end{proof}

Theorem~\ref{thm:kempe-ordering} shows that the Kempe invariant affects the fidelity gain provided by a controller in quantum teleportation. For pure GME states with fixed pairwise concurrences and three-tangle, the optimized assisted fidelities remain unchanged, whereas the unassisted standard fidelities decrease strictly with increasing Kempe invariant. Consequently, the control power increases for every choice of controller. This dependence distinguishes $e_{\mathrm{GCP}}$ from the GME concurrence $C_{\mathrm{GME}}$ \cite{Ma2011} and the teleportation-capability measures $\mathcal T_{\min}$ and $\mathcal T_{\mathrm{GM}}$ \cite{Choi2023}, all of which are determined by the pairwise concurrences and three-tangle and therefore remain unchanged under these constraints.

\begin{figure}[tb]
\centering
\includegraphics[width=\columnwidth]{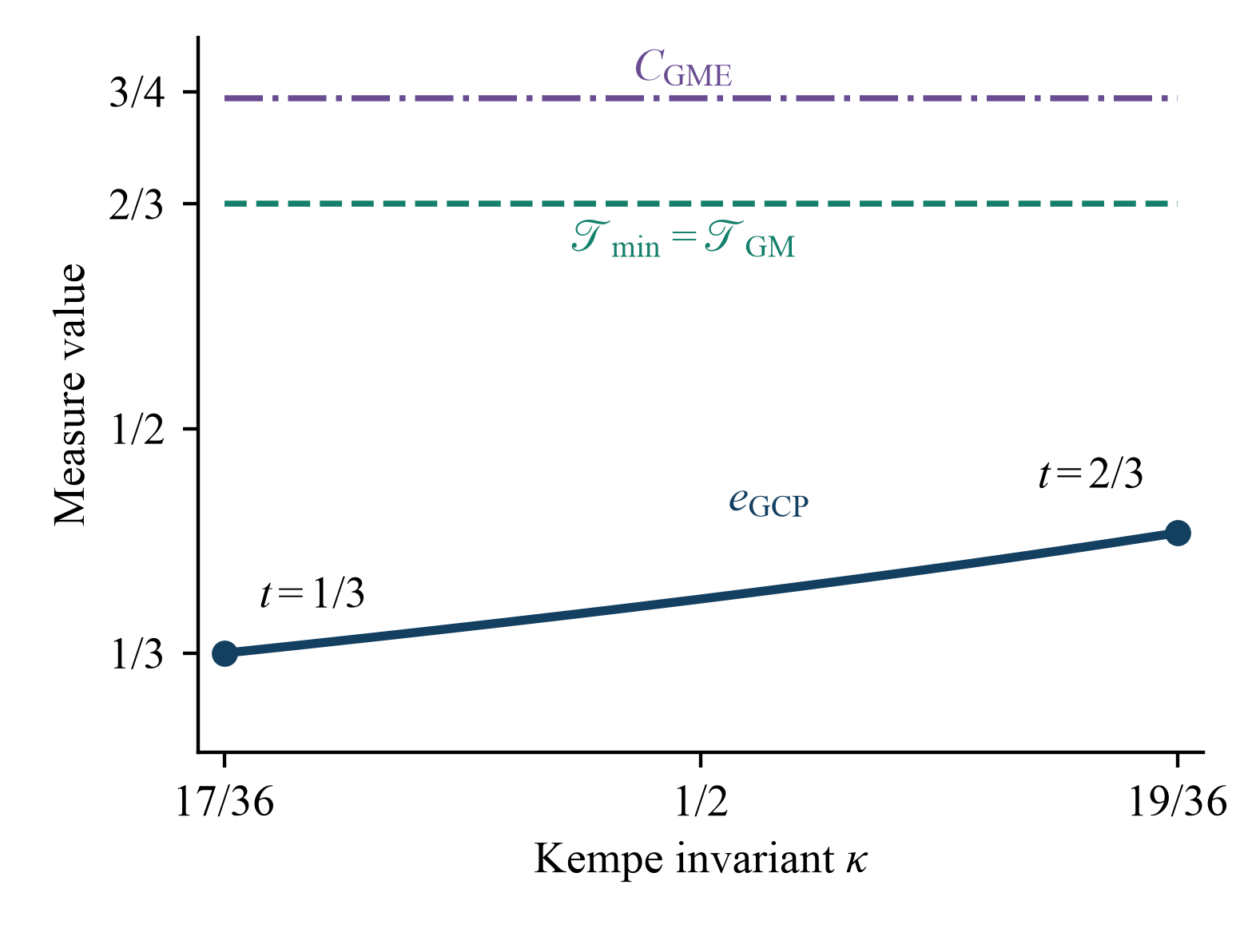}
\caption{Kempe dependence for the continuous family in Eq.~(\ref{eq:kempe-continuous-family}), with fixed $C_{AB}=C_{AC}=C_{BC}=\tau=1/3$. The measure $e_{\mathrm{GCP}}$ increases strictly with $\kappa$, as established by Theorem~\ref{thm:kempe-ordering}. The GME concurrence $C_{\mathrm{GME}}$ remains constant at $\sqrt5/3$, while $\mathcal T_{\min}=\mathcal T_{\mathrm{GM}}=2/3$ share a single horizontal curve. The markers indicate $t=1/3$ and $t=2/3$, where $e_{\mathrm{GCP}}=1/3$ and $1-1/\sqrt3$, respectively.}
\label{fig:kempe-fixed-data}
\end{figure}

To illustrate this ordering, consider the continuous family in the Ac\'in canonical form, a representation of three-qubit pure states up to local unitaries \cite{Acin2000},
\begin{equation}
\begin{array}{r@{}l}
\ket{\psi(t)}&{}=\lambda_0\ket{000}+\lambda_1e^{i\varphi}\ket{100}\\[3pt]
&{}\quad+\lambda_2\ket{101}+\lambda_3\ket{110}+\lambda_4\ket{111},
\end{array}
\label{eq:kempe-continuous-family}
\end{equation}
where $1/3\le t\le2/3$ and
\[
\begin{array}{c}
\lambda_0=\sqrt t,\qquad
\lambda_1=\sqrt{1-t-\frac{5}{36t}},\\[6pt]
\lambda_2=\lambda_3=\frac{1}{6\sqrt t},\qquad
\lambda_4=\frac{1}{2\sqrt{3t}},\\[6pt]
\displaystyle\cos\varphi=
\frac{54t-72t^2-7}{\sqrt{3(36t-36t^2-5)}},\qquad 0\le\varphi\le\pi.
\end{array}
\]
Throughout this interval, the states have fixed concurrence--tangle data,
$C_{AB}=C_{AC}=C_{BC}=\tau=1/3$, while their Kempe invariant is
\[
\kappa(t)=\frac14+\frac t3+\frac{1}{27t}.
\]
It increases strictly from $17/36$ to $19/36$. Consequently, Theorem~\ref{thm:kempe-ordering} gives a strict increase of $e_{\mathrm{GCP}}$ along the family. In contrast, $C_{\mathrm{GME}}=\sqrt5/3$ and $\mathcal T_{\min}=\mathcal T_{\mathrm{GM}}=2/3$ remain constant, as shown in Fig.~\ref{fig:kempe-fixed-data}.

The two endpoints make the fidelity interpretation explicit. For every controller, the optimized assisted fidelity is $F_{\mathrm{CT}}^X=8/9$ throughout the family. At $t=1/3$ and $t=2/3$, the unassisted standard fidelities are $7/9$ and $(5+\sqrt3)/9$, respectively, giving
\[
e_{\mathrm{GCP}}(\ket{\psi(1/3)})=\frac13,\qquad
e_{\mathrm{GCP}}(\ket{\psi(2/3)})=1-\frac1{\sqrt3}.
\]
These values follow from the largest roots of Eq.~(\ref{eq:kempe-characteristic-polynomial}). Thus, the increase in $e_{\mathrm{GCP}}$ reflects a larger fidelity gain supplied by each controller as the unassisted teleportation fidelity decreases.

\section{Conclusion}
\label{sec: Conclusion}

We have introduced a GME measure for three-qubit states based on the fidelity gain supplied by a controller in quantum teleportation. For pure states, $e_{\mathrm{GCP}}$ is the geometric mean of the three normalized control powers, with the controller's measurement optimized independently for each choice of controller. We proved that it vanishes exactly on biseparable states and is nonincreasing on average under LOCC. Its convex-roof extension retains these properties and is convex, giving a GME measure for mixed three-qubit states. The pure-state maximum is unity, attained uniquely by the standard GHZ state up to local unitaries. Within the W class, the corresponding maximum is $1/3$, attained uniquely by the standard W state up to local unitaries.

We further established that, for genuinely entangled pure states with fixed pairwise concurrences and three-tangle, $e_{\mathrm{GCP}}$ increases strictly with the Kempe invariant. Under these constraints, a larger Kempe invariant lowers the unassisted standard-teleportation fidelity for every pair while leaving the optimal assisted fidelities unchanged. This dependence distinguishes our measure from the GME concurrence \cite{Ma2011} and existing teleportation-capability measures \cite{Choi2023}, which remain constant under the same constraints. The resulting increase in the fidelity gain supplied by every controller gives the Kempe-dependent ordering an operational interpretation, as illustrated by a continuous family of pure states.

This constrained ordering is also shared by the entanglement monotone $\omega$ introduced by Szalay \cite{Szalay2026}, although the two measures need not agree over the full pure-state space. For example, $e_{\mathrm{GCP}}$ ranks $\ket{G_{\pi/12}}$ above the standard W state, whereas $\omega$ ranks it below. The common ordering thus admits an interpretation in terms of the benefit supplied by a controller in teleportation. The underlying FEF dependence also connects our result to the concurrence dynamics studied by Song {\it et al}. \cite{Song2026b}. For pure three-qubit inputs with the same initial concurrence–tangle data, their result orders pairwise concurrence under identical, independent local depolarization. The initial concurrences coincide, but their decay and entanglement robustness can differ, with smaller initial Kempe invariants giving no smaller concurrence throughout the evolution. In our result, smaller Kempe invariants already correspond to larger FEFs of the initial reduced states and hence higher standard-teleportation fidelities, without introducing a noise model. Identical local depolarization reduces these FEF differences while preserving their strict ordering until complete depolarization. The operational distinction therefore exists before noise is applied, whereas the noisy-concurrence ordering describes separate dynamical information.

For mixed states, obtaining computable bounds on the convex-roof measure and determining its relation to the fidelity gain available directly from a mixed resource remain open problems. Such bounds would help assess GME in noisy implementations of controlled teleportation. Analyzing the measure under local noise would also allow resource states to be compared by the GME they retain. Extensions to more parties or higher local dimensions require specifying the assisted teleportation protocol and identifying combinations of control powers that preserve faithfulness and average LOCC monotonicity.


\section*{AI Statement}
ChatGPT (OpenAI), including GPT-6.1 Sol and GPT-6.1 Astra, was used for language editing, LaTeX assistance, and mathematical discussion related to the proofs of
Lemma~\ref{lem:gcp-monotonicity} and Theorem 2 and the construction of states in Eq.~(\ref{eq:kempe-continuous-family}). The authors independently verified the mathematical suggestions and take full responsibility for the final manuscript
\begin{acknowledgments}
S.L. acknowledges support from the Institute of Information \& Communications Technology Planning \& Evaluation (IITP) grant funded by the Ministry of Science and ICT (MSIT) (No. RS-2025-02304540).
M.C. acknowledges support from the Korea Institute of Science and Technology Information (KISTI) R\&D (Grant No. K26L3M3C3).
J.L. acknowledges support from the Korea Institute of Science and Technology (26E0001).
S-W.L. acknowledges support from the National Research Foundation of Korea (Grants No. RS-2022-NR068812).
\end{acknowledgments}

\appendix

\section{Proof of Lemma~\ref{lem:gcp-filtering}}
\label{app:fef-filtering}

\begin{proof}
For a factorization $\rho_{YZ}=LL^\dagger$, the two-qubit CoA formula reads $C_a(\rho_{YZ})=\|L^T(\sigma_y\otimes\sigma_y)L\|_1$ \cite{Laustsen2003}, where $\|\cdot\|_1$ is the trace norm. For $p_t>0$, a factorization of $\rho_{YZ}^{(t)}$ is obtained by replacing $L$ with $(K_t\otimes I)L/\sqrt{p_t}$. Using $K_t^T\sigma_yK_t=(\det K_t)\sigma_y$ then gives
\[
p_tC_a(\rho_{YZ}^{(t)})=|\det K_t|C_a(\rho_{YZ}).
\]
If $p_t=0$, then $K_t$ must be singular, so its contribution to $s$ also vanishes. Summing over the nonzero-probability outcomes proves the first relation in Eq.~(\ref{eq:gcp-filtering}). The singular values of each $K_t$ also give
\[
s=\sum_t|\det K_t|\le\frac12\sum_t\Tr(K_t^\dagger K_t)=1.
\]

For the FEF bound, extend the definition homogeneously to every positive semidefinite operator $Q$ by
\[
f(Q)=\max_{\ket{e}_{YZ}}\bra{e}Q\ket{e}_{YZ},
\]
where the maximum is over normalized maximally entangled states. Thus $f(\sigma_t)=p_tf(\rho_{YZ}^{(t)})$ for $p_t>0$, and $f(0)=0$.

Choose a maximizing vector $\ket{e_*}_{YZ}$ for $\rho_{YZ}$ and define the unnormalized vector and its normalized version by
\[
v=\frac{\rho_{YZ}\ket{e_*}_{YZ}}{\sqrt{f(\rho_{YZ})}},
\qquad q=\|v\|^2,
\qquad \ket v=\frac{v}{\sqrt q}.
\]
The maximum exists by compactness, and $f(\rho_{YZ})\ge1/4$ because the four Bell-state overlaps sum to one. For any vector $w$, the Cauchy--Schwarz inequality applied to $\sqrt{\rho_{YZ}}\,w$ and $\sqrt{\rho_{YZ}}\ket{e_*}_{YZ}$ gives
\[
w^\dagger vv^\dagger w
=\frac{|w^\dagger\rho_{YZ}\ket{e_*}_{YZ}|^2}{f(\rho_{YZ})}
\le w^\dagger\rho_{YZ}w.
\]
Thus
\begin{equation}
vv^\dagger\le\rho_{YZ}.
\label{eq:fef-component-bound}
\end{equation}
Taking the expectation in $\ket{e_*}_{YZ}$ gives
\[
f(\rho_{YZ})=|\bra{e_*}_{YZ}v|^2\le f(vv^\dagger)\le f(\rho_{YZ}),
\]
so $f(vv^\dagger)=f(\rho_{YZ})=qf(\ket v)$.

Since $\rho_{YZ}-vv^\dagger\ge0$, its spectral decomposition gives the pure-state ensemble
\[
\rho_{YZ}=q\ket v\bra v+\sum_jr_j\ket{\eta_j}\bra{\eta_j},
\qquad q+\sum_jr_j=1.
\]
where $r_j\ge0$ and $\ket{\eta_j}$ are the normalized eigenvectors of the remainder. By the definition of concurrence of assistance,
\[
qC(\ket v)\le C_a(\rho_{YZ}).
\]

For $v_t=(K_t\otimes I)v$, define $q_t=\|v_t\|^2$ and $\ket{v_t}=v_t/\sqrt{q_t}$ when $q_t>0$. Applying the filter $K_t\otimes I$ to Eq.~(\ref{eq:fef-component-bound}) gives $v_tv_t^\dagger\le\sigma_t$, and completeness gives $\sum_tq_t=q$. If $M$ is the coefficient matrix of $\ket v$, then the coefficient matrix of $\ket{v_t}$ is $\sqrt{q/q_t}K_tM$. Thus
\[
q_tC(\ket{v_t})=q|\det K_t|C(\ket v).
\]

Using $f(\ket{v_t})=[1+C(\ket{v_t})]/2$ for normalized pure two-qubit states \cite{Grondalski2002}, we obtain
\[
\begin{array}{r@{}l}
\displaystyle\sum_t p_tf(\rho_{YZ}^{(t)})
&\displaystyle{}=\sum_tf(\sigma_t)\ge\sum_tq_tf(\ket{v_t})\\[3pt]
&\displaystyle{}=\frac q2+\frac{qs}{2}C(\ket v)\\[3pt]
&\displaystyle{}=f(\rho_{YZ})-\frac{1-s}{2}qC(\ket v)\\[3pt]
&\displaystyle{}\ge f(\rho_{YZ})-\frac{1-s}{2}C_a(\rho_{YZ}),
\end{array}
\]
where the last inequality uses $s\le1$ and $qC(\ket v)\le C_a(\rho_{YZ})$.

\end{proof}

\section{Characteristic-polynomial comparison}
\label{app:kempe-polynomial}

We derive the coefficients in Eq.~(\ref{eq:kempe-characteristic-polynomial}). For a pure three-qubit state, set $\rho=\rho_{BC}$ and $R=(\rho+\widetilde\rho)/2$, with local purities $\alpha$, $\beta$, and $\gamma$ as defined in the proof of Theorem~\ref{thm:kempe-ordering}.

The two-qubit spin-flip identity \cite{Rungta2001} is
\begin{equation}
\widetilde\rho=\rho+I_4-\rho_B\otimes I_2-I_2\otimes\rho_C.
\label{eq:spinflip-inversion}
\end{equation}
It follows by applying
$\sigma_y M^T\sigma_y=(\Tr M)I_2-M$ to the two tensor factors.
Global purity gives $\Tr\rho^2=\alpha$ and
$\Tr\rho^3=\Tr\rho_A^3$.

For the second moment, Eq.~(\ref{eq:spinflip-inversion}) gives
$\Tr(\rho\widetilde\rho)=\alpha+1-\beta-\gamma$.
Since $\rho$ and $\widetilde\rho$ have the same spectrum,
\[
\Tr R^2=\frac12\bigl[\Tr\rho^2+\Tr(\rho\widetilde\rho)\bigr]
=\alpha+\frac{1-\beta-\gamma}{2}.
\]

For the third moment, the Schmidt identity
$(\rho_A\otimes I_{BC})\ket\psi=(I_A\otimes\rho_{BC})\ket\psi$
implies
\[
\begin{array}{r@{}l}
\Tr[\rho^2(\rho_B\otimes I_2)]&{}=\Tr[(\rho_A\otimes\rho_B)\rho_{AB}],\\[3pt]
\Tr[\rho^2(I_2\otimes\rho_C)]&{}=\Tr[(\rho_A\otimes\rho_C)\rho_{AC}].
\end{array}
\]
Using Eq.~(\ref{eq:spinflip-inversion}) and the permutation-symmetric
Kempe invariant in Eq.~(\ref{eq:kempe-invariant}), we obtain
\[
\Tr(\rho^2\widetilde\rho)
=\alpha+\frac{\Tr\rho_A^3-\Tr\rho_B^3-\Tr\rho_C^3-2\kappa}{3}.
\]
Spin-flip symmetry gives
$\Tr(\rho\widetilde\rho^{\,2})=\Tr(\rho^2\widetilde\rho)$.
Expanding $\Tr R^3$ and using
$\Tr\rho_X^3=(3\Tr\rho_X^2-1)/2$ for each qubit yields
\[
\begin{array}{r@{}l}
\Tr R^3&\displaystyle{}=\frac{\Tr\rho_A^3+3\Tr(\rho^2\widetilde\rho)}{4}\\[5pt]
&\displaystyle{}=\frac38(4\alpha-\beta-\gamma)-\frac\kappa2.
\end{array}
\]

To express $\det R$ in terms of the local spectra, we first relate the determinant of $\rho+\widetilde\rho$ to that of $\rho-\widetilde\rho$.
For any $4\times4$ matrices $A,B$ of rank at most two,
multilinearity in the columns gives $\det(A+tB)=d\,t^2$ for some $d$ because
every term using three columns from either matrix vanishes. Therefore
$\det(A+B)=\det(A-B)$, including rank-deficient cases.
Here both $\rho$ and $\widetilde\rho$ have rank at most two, so
\[
\det R=\frac1{16}\det(\rho+\widetilde\rho)
=\frac1{16}\det(\rho-\widetilde\rho).
\]
By Eq.~(\ref{eq:spinflip-inversion}), the last matrix is
$\rho_B\otimes I_2+I_2\otimes\rho_C-I_4$.
If the local Bloch-vector lengths are $b,c$, its four eigenvalues are
$(\pm b\pm c)/2$. Thus
\[
\det R=\frac{(b^2-c^2)^2}{256}
=\frac{(\beta-\gamma)^2}{64},
\qquad \beta=\frac{1+b^2}{2},\quad\gamma=\frac{1+c^2}{2}.
\]
Finally, since $\Tr R=1$, Newton's identities give
\[
\begin{array}{r@{}l}
\det(uI_4-R)&\displaystyle{}=u^4-u^3+\frac{1-\Tr R^2}{2}u^2\\[5pt]
&\displaystyle{}-\frac{1-3\Tr R^2+2\Tr R^3}{6}u+\det R.
\end{array}
\]
Substituting the moments and determinant above yields Eq.~(\ref{eq:kempe-characteristic-polynomial}), and hence Eq.~(\ref{eq:kempe-polynomial-difference}).

\bibliography{references}
\end{document}